\documentclass{article}
\usepackage{amssymb,epsfig}
\begin{document}
\centerline{\large\bf SELF--SIMILAR AND CHARGED RADIATING SPHERES:}
\vspace*{0.035truein}
\centerline{\large\bf AN ANISOTROPIC APPROACH}
\vspace*{0.245truein}
\centerline{\footnotesize{\large W. Barreto\footnote{Departamento de F\'\i isica, 
Centro de F\'\i sica Fundamental, Facultad de Ciencias, Universidad de Los Andes}
 B. Rodr\'\i guez\footnote{Computational Science Research Center,
College of Sciences,
San Diego State University, San Diego, California.}
L. Rosales\footnote{Laboratorio de F\'\i sica Computacional,
Universidad Experimental Polit\'ecnica ``Antonio Jos\'e de
Sucre'', Puerto Ordaz, Venezuela.}
and O. Serrano\footnote{Departamento de Ciencia y Tecnolog\'\i a,
 Universidad Experimental de Guayana,
 Puerto Ordaz, Venezuela.}}}
\baselineskip=12pt
\vspace*{0.21truein}
\date{\today}

\begin{abstract}
Considering charged fluid spheres as anisotropic sources
and the diffusion limit as the transport mechanism, we
suppose that the inner space--time admits self--similarity. 
Matching the interior solution with the Reissner--Nordstr\"om--Vaidya exterior one,
we find an extremely compact and oscillatory final state with a redistribution of 
the electric charge function and non zero pressure profiles.  

\vspace*{0.21truein}

\end{abstract}
\section{Introduction}
It is well known that astrophysical objects are not significantly charged,
For this reason charged bodies have limited interest 
(for an historical overview see Section I in Ref. \cite{remlz03}).
Observed stars, like the Sun or a neutron star, cannot support a great
amount of charge. This is so basically because the particles that compose
a star have a huge charge to mass ratio, as is the case for a proton or an
electron. For cold stars the electric charge is about $100$ Coulombs per solar mass \cite{g00}.
Nevertheless, in compact stars the equilibrium conditions are different,
allowing some more net charge \cite{b71}. 
For highly relativistic stars near to form a black hole, the huge gravitational pull
can be balanced by huge amounts of net charge. On this kind of configurations
we are concerned in this paper and have been considered by other authors 
\cite{remlz03},\cite{b71}--\cite{ar01}.

To include the electric charge a  number of authors
make additional assumptions such as an equation of state or a
relationship between metric variables \cite{hs97}--\cite{hm84}.
 Bonnor and Wickramasuriya \cite{bw75} have studied
electrically charged matter, with electrostatic
repulsion balancing the gravitational attraction.
Most works have been done under static conditions,
including the ones by Ivanov \cite{i02}, who exhaustively surveyed
static charged perfect fluid spheres in general relativity, and Ray {\it et
al.}  \cite{remlz03}, who studied the effect of electric charge in compact stars
and its consequences on the gravitational collapse.
Bekenstein \cite{b71}
found that for highly relativistic stars, whose radius is on the verge
of forming an event horizon, the large gravitational pull can be
balanced by large amounts of net charge.

In this paper we study an electrically charged matter distribution 
as an anisotropic fluid. 
It is well known that different energy--momentum tensors can lead to
the same space--time \cite{carotplus}. For instance, under spherical symmetry
viscosity can be considered as a special case of anisotropy \cite{brt}.
To illustrate our approach to obtain
dynamical models we consider the diffusion approximation as the transport mechanism, and a
self--similar space--time for the inner region. Our treatment allows
different scenarios for the gravitational collapse, including the 
reported some years ago \cite{bd99}. In particular we are now interested in
non stationary initial conditions to explore the fate of the gravitational collapse.
The set problem has some physical and geometrical features that we review briefly, these are: anisotropy,
dissipation and self--similarity.

For certain density ranges, local
anisotropy in pressure can be physically justified in self--gravitating
systems, since different kinds of physical phenomena may take place,
giving rise to local anisotropy and in turn relaxing the upper limits
imposed on the maximum value of the surface gravitational potential \cite{l33}.
The influence of local anisotropy in general relativity has been
studied mostly under static conditions (see \cite{hs97} and references therein). 
Herrera {\it et al.} \cite{hetal04} have reported a
general study for spherically symmetric dissipative anisotropic fluids
with emphasis on the relationship among the Weyl tensor, the shear
tensor, the anisotropy of the pressure and the density inhomogeneity.
In another context, a scalar field with non--zero spatial gradient is
an example of a physical system where the pressure is clearly
anisotropic. 
Boson stars, hypothetical
self-gravitating compact objects resulting from the coupling of a
complex scalar field to gravity, are systems where anisotropic
pressure occurs naturally. Similarly, the energy--momentum tensor of
both electromagnetic and fermionic fields are anisotropic \cite{dg02}.
We want to stress that in order to have isotropy we need an extra
assumption on the behavior of the fields or on the fluid describing
interiors.

The proposed dissipative process 
is suggested by the fact that they provide the only plausible
mechanism to carry away the bulk of binding energy, leading to a
neutron star or black hole \cite{ks79}. Furthermore, in cores of densities
close to $10^{12}\, g/cm^3$ the mean free path for neutrinos becomes small
enough to justify the use of the diffusion approximation \cite{a77,k78}. 
 Here we do not discuss the temperature distribution during diffusion; we consider
this transport mechanism because it provides an efficient way to
radiate energy. Additional studies are necessary to explore the
consequences of incorporating a hyperbolic theory of dissipation to
overcome the difficulties inherent to parabolic theories.

Few exact solutions to the Einstein--Maxwell equations are relevant
to gravitational collapse. For this reason, new collapse solutions
are very useful, even if they are simplified ones \cite{op90}. It
is well known that the field equations admit homothetic motion
\cite{ct71}--\cite{lz90}. Applications of self--similarity
range from modeling black holes to producing counterexamples
to the cosmic censorship conjecture \cite{ch74}--\cite{cc98}.
It is well established that in the critical
gravitational collapse of an scalar field the space--time can be self--similar
\cite{c93}--\cite{g99}.  We can expect on physical grounds that this scenario
remains similar for charged matter \cite{bcgn02}. 

In this work the Darmois--Lichnerowicz junction conditions
at the surface of the distribution are satisfied. We match the
self--similar interior solution with the Reissner--Nordstr\"om--Vaidya exterior one.
We obtain a model that resembles the Seidel--Suen
solitonic star formed by a real and massive scalar field \cite{ss91}.

Section \ref{main_eq} contains the field equations, written in such form that the
electrically charged fluid  can be considered anisotropic. 
In this section we also discuss the junction
conditions to match the dissipative interior space--time with the
exterior space--time, a Reissner--Nordstr\"om--Vaidya one, and write the
generalized Tolman--Oppenheimer--Volkoff equation. Section \ref{selfsimilar} contains a
review of self--similarity for the sake of completeness. Then we propose
a simple solution to include electric charge by means of the active
mass. We show two example solutions in section \ref{modeling} 
and finally discuss our results 
in section \ref{conclusion}.

\section{The field equations and Junction conditions}\label{main_eq}
Let us consider a non static distribution of matter which is spherically
symmetric and consists of charged fluid of energy density $\rho$, pressure $p$,
electric energy density $\epsilon$ and radiation energy flux $q$ diffusing in
the radial direction, as measured by a local Minkowskian observer comoving
with the fluid (with velocity $-\omega$).
 In radiation coordinates \cite{b64} the metric takes the form
\begin{equation}
ds^2=e^{2\beta}\bigg(\frac Vrdu^2+2du\,dr\bigg)
-r^2\bigg(d\theta ^2+\sin^2\theta\,d\phi ^2\bigg), 
\end{equation} 
where $\beta$ and $V$ are functions of $u$ and $r$. 
As it is well known for the spherical symmetry $F^{ur}=-F^{ru}$ are the only
non vanishing electromagnetic field tensor components. Now, defining the
function $C(u,r)$ by the relation
\begin{equation}
F^{ur}=\frac{C}{r^2}e^{-2\beta}
\end{equation}
the inhomogeneous Maxwell equations become
\begin{equation}
C_{,r}=4\pi r^2J^ue^{2\beta},
\end{equation}
and
\begin{equation}
C_{,u}=-4\pi r^2J^re^{2\beta},
\end{equation}
where the comma subscript represents partial derivative with respect to the
indicated coordinate; $J^u$ and $J^r$ are the temporal and radial components of
the electric current four vector, respectively. Thus, the function $C(u,r)$ is
naturally interpreted as the charge within the radius $r$ at the time $u$. 
Thus, the inhomogeneous Maxwell field equations entail the conservation of charge
\begin{equation}
U^\mu C_{,\mu}=0,
\end{equation}
where the four--velocity is given by
\begin{equation}
U^\mu=e^{-\beta}\left[\sqrt{\frac{r}{V}}\left(\frac{1-\omega}{1+\omega}\right)\delta^\mu_u+
\sqrt{\frac{V}{r}}\frac{\omega}{(1-\omega^2)^{1/2}}\delta^\mu_r\right].
\end{equation}
We can write the conservation equation in a more suitable
form, that is
\begin{equation}
C_{,u}+\frac{dr}{du} C_{,r}=0,
\label{ce}
\end{equation}
where 
\begin{equation}
\frac{dr}{du}=e^{2\beta}\frac{V}{r}\frac{\omega}{1-\omega}\label{eq:drdu}
\end{equation}
is the matter velocity.
It is clear that inside a sphere comoving with the fluid the
charge is conserved. 

Introducing the mass function by means of
\begin{equation}
\mu=\frac{1}{2}\left(r-Ve^{-2\beta}\right),
\end{equation}
we can write the Einstein field equations as follows
\begin{equation}
\frac{\hat\rho + p_r\omega^2}{1-\omega^2}+\frac{2\omega q}{1-\omega^2}=
-\frac{e^{-2\beta}\mu_{,u}}{4\pi r(r-2\mu)} +\frac{\mu_{,r}}{4\pi r^2}
\label{eq:one}
\end{equation}
\begin{equation}
\bar \rho=\frac{\mu_{,r}}{4\pi r^2}
\end{equation}
\begin{equation}
\bar\rho + \bar p=\frac{\beta_{,r}}{2\pi r^2}(r-2\mu)
\end{equation}  
\begin{eqnarray}
p_t=-\frac{1}{4\pi}\beta_{,ur}e^{-2\beta}+\frac{1}{8\pi}(1-2\mu/r)
(2\beta_{,rr}+4\beta_{,r}^2-\beta_{,r}/r)\nonumber \\
+\frac{1}{8\pi r}[3\beta_{,r}
(1-2\mu_{,r})-\mu_{,rr}]
\end{eqnarray}
where 
$$
\bar p=\frac{p_r-\omega\hat\rho}{1+\omega}-\frac{1-\omega}{1+\omega}q,
$$
and
$$
\bar \rho=\frac{\hat\rho-\omega p_r}{1+\omega}-\frac{1-\omega}{1+\omega}q,
$$
with $\hat \rho=\rho + \epsilon$, $p_r=p-\epsilon$, $p_t=p+\epsilon$
 and the electric energy density
$\epsilon=E^2/8\pi$, where $E=C/r^2$ is the local electric field intensity.
Observe that we have introduced the mass function $\mu$ instead the usual
total mass $\tilde m=(r-Ve^{-2\beta})/2+C^2/2r$. 

If we define the degree of local anisotropy induced by charge as
$\Delta=p_t-p_r=2\epsilon$, the electric charge determines such a degree at any point.

It can be shown that the Tolman--Oppenheimer--Volkoff equation reduces to
\begin{eqnarray}
\frac{\partial\bar p}{\partial r}+\frac{\bar\rho+\bar p}{1-2\mu/r}
\bigg[4\pi r \bar p +\mu/r^2\bigg]
-e^{-2\beta}\bigg(\frac{\bar\rho+\bar p}{1-2\mu/r}\bigg)_{,u}=\nonumber \\
\frac{2}{r}
\bigg(p_t-\bar p\bigg).
\label{eq:tov}
\end{eqnarray}
This generalized equation is the same as for an anisotropic fluid \cite{cosenzaetal}.

The most general junction conditions for our
system can be written
as \cite{bd99} 
\begin{equation}
\mu_a= \tilde m_a-\frac{C_T^2}{2a},
\end{equation}
\begin{equation}
\beta_a=0,
\end{equation}
and 
\begin{equation}
-\beta _{,u}|_a+(1-2\mu _{a}/a)\beta _{,r}|_a-\frac{\mu _{,r}|_a}{2a}+
\frac{C_{T}^{2}}{4a^{3}}=0,
\end{equation}
where a subscript $a$ indicates that the quantity is evaluated at the surface, $\tilde m_a$  and
$C_T$ are the total mass and the total charge, respectively. 
Remarkably, the last equation is equivalent to
\begin{equation}
p_{a}=q_{a}.
\end{equation}

We have obtained a system of equations which are equivalent to the anisotropic
matter case (see \cite{hs97} and references therein). In the next
section we will see how this approach lead us to an extra
simplification.

Up to this point all the written equations are general. We can try to solve the system using
numerical techniques or a seminumerical approach to gain
some insight. The last method is worked out in the next section.

\section{The self--similar space--time: A simple solution}\label{selfsimilar}
We shall assume that the spherical
distribution admits a one--parameter group of homothetic motion.
A homothetic vector field on the manifold is one that satisfies
$\pounds_\xi {\bf g}=$2$n{\bf g}$ on a local chart,
 where $n$ is a constant on the
manifold and $\pounds$ denotes the Lie derivative operator. If $n \ne 0$ we
have a proper homothetic vector field and it can always be scaled so as to have
$n = 1$; if $n = 0$ then $\xi$ is a Killing vector on the manifold.
So, for a constant rescaling, $\xi$ satisfies
\begin{equation}
\pounds_\xi{\bf g}=2{\bf g}
\end{equation}
and has the form
\begin{equation}
\xi =\Lambda (u,r)\partial_u  +\lambda (u,r)\partial_r.
\end{equation}  
The homothetic
 equations reduce to
\begin{equation}
\xi(\chi)=0,\label{se1}
\end{equation}
\begin{equation}
\xi(\upsilon)=0,\label{se2}
\end{equation}
where $\lambda=r$, $\Lambda=\Lambda(u)$,
$\chi=\mu/r$ and  $\upsilon=\Lambda e^{2\beta}/r$.
Therefore, $\chi=\chi(\zeta)$ and $\upsilon=\upsilon(\zeta)$ are solutions
 if the self--similar
 variable is defined as
\begin{equation}
\zeta\equiv r\, e^{- \int du/\Lambda}.
\end{equation}
We have reported \cite{bd99,bpr99} a 
simple homothetic solution
 which can be written
for the present case as
\begin{equation}
\mu=\mu_a(u) (r/a)^{k+1}
\end{equation}
and
\begin{equation}
e^{2\beta}=(r/a)^{l+1}
\end{equation}
with $k$ and $l$ constants. It can be shown that for a given
$k$, $l$ must be a root of a complicated seventh degree 
polynomial in order to
satisfy the additional symmetry equations.
It is
obvious that these solutions satisfy the continuity of
the first fundamental form, and to satisfy the continuity
of the second fundamental form, the radial velocity at the
surface must be
\begin{equation}
\omega_a=1-\frac{2(l+1)(1-2\mu_a/a)}{2(k+1)\mu_a/a-C_T^2/a^2}.
\label{omega}
\end{equation}
\subsection{Surface equations}
Now, to completely determine the metric we have two
differential equations at the surface, the equation
(\ref{eq:drdu}) and the field equation (\ref{eq:one}) which
evaluated at the surface become respectively
\begin{equation}
\dot a=(1-2\mu_a/a)\frac{\omega_a}{1-\omega_a}\label{ap}
\end{equation}
and
\begin{equation}
\dot\mu_a=-Q(1-2\mu_a/a)+\dot a \frac{C_T^2}{2a^2}, \label{mup}
\end{equation}
where a dot over a variable denotes the  derivative with respect to time.
The quantity $Q$ is defined as
\begin{equation}
Q\equiv 4\pi a^2 q_a \left(\frac{1+\omega_a}{1-\omega_a}\right),
\label{Q}
\end{equation}
 and can be explicitly determined using equation (\ref{eq:tov})
evaluated at the surface, resulting in
\begin{eqnarray}
Q=\{(1-2\mu_a/a)[2(l+1)^2+k(k+1+3(l+1))]-2(C_T/a)^2 \nonumber \\
-k(k+1+3(l+1))\} 
\{k+1-(1-2\mu_a/a)(k+l+2)-(C_T/a)^2\}/
\nonumber \\
\left[4(1-2\mu_a/a)(l+1)a^2\right].\label{Q+}
\end{eqnarray}

The surface equations (\ref{ap}) and (\ref{mup}) together with
equations (\ref{omega}) and (\ref{Q+}) can be integrated numerically
(using Runge--Kutta for instance) for initial conditions
$a(0)$ and $\mu_a(0)$ and some $k$ and $C_T$. 
The only restrictions on these conditions and parameters come from the need to
get a physically acceptable model.

For $k=0$ (which implies $l=0$) we obtain the same results of reference \cite{bd99}. 
As we have mention before, we gain some simplification with the anisotropic approach
of this paper, that is, one of the similarity equations
(Eq. (23) in Ref. \cite{bd99}) was decoupled, 
leading us to less restrictive models.

\subsection{Integrating the conservation equation}
Once the surface equations are integrated we have to integrate
the conservation equation (\ref{ce}) to obtain all the physical
variables inside the source. To do that we use comoving spatial markers
$x=r/a$. Thus the conservation equation can be written as
\begin{equation}
C_{,u}=-\frac{dx}{du} C_{,x}
\end{equation}
which is a wave--like equation that can be integrated numerically
using the Lax method (with the appropriate Courant--Friedrichs--Levy (CFL)
condition). The evolution of the conservation equation is restricted
by the surface evolution and was implemented as follows
\begin{equation}
C^{n+1}_j=\frac{1}{2}\left(C^n_{j+1}+C^n_{j-1}\right)+
\frac{\delta u}{2\delta x} \left(\frac{dx}{du}\right)^n_j
\left(C^n_{j+1}-C^n_{j-1}\right).
\end{equation}
The superscript $n$ indicates the hypersurface $u=n\delta u$
and the subscript $j$ the spatial position for a comoving
observer at $x=j\delta x$. We typically used $\delta u=10^{-2}$
with a CFL condition $\delta u=2\delta x$. In order to integrate
the conservation equation we need to specify a boundary
condition and an initial condition.
For this work we have used the boundary condition
\begin{equation}
C(x=0,u) = 0,
\end{equation}  
with an initial condition
\begin{equation}
C(x,u=0) = C_T\, x^{\mathcal P},
\end{equation}
where the power ${\mathcal P}$ allows us 
to test the sensitivity to the initial conditions of our results.

\section{Modeling}\label{modeling}
\subsection{Evolving towards dust: $k=0$}\label{A}
As a test of our approach to dealing with electric charge as anisotropy is to
reproduce the results reported in \cite{bd99}. For this scenario 
$k=l=0$, $a(0)=3.5$, $\mu_a(0)=1$, ${\mathcal P}=1$ and $C_T=0.09$.
 Even when the initial mass $m_a (0)$ is not equal to one, we obtain the same
results, that is, the collapse is halted and the boundary oscillates
when the total electric charge asymptotically approximates the total
mass $m_a$ and the radius equals asymptotically to twice the
 total electric charge. The ratio $P/\rho$ as a function of time is the same for any region.
 Also, as the distribution evolves,
becomes dust asymptotically with damped oscillations and the pressure
at the surface coincides with the heat flow. The matter velocity profiles
at all regions have damped oscillations that evolve to the static regime.
A striking result not reported before is the ability to obtain any interior profile for
matter velocity or electric charge function using the simple rule
\begin{equation}
{\mathcal F}(u,x)=x{\mathcal F}(u,1),
\end{equation}
where ${\mathcal F}$ could be either $dr/du$ or $C$, that is,
 once the surface profile is determined we can go without additional work inside the
distribution. Another interesting result is that we can 
get the interior profiles for pressure and energy density by
\begin{equation}
{\mathcal G}(u,x)={\mathcal G}(u,1)/x^2,
\end{equation}
where ${\mathcal G}$ could be $\rho$ or $p$. 

\subsection{Redistribution of the electric charge: $k=0.1$}\label{B}
\begin{figure}
\centerline{\epsfxsize=4.in\epsfbox{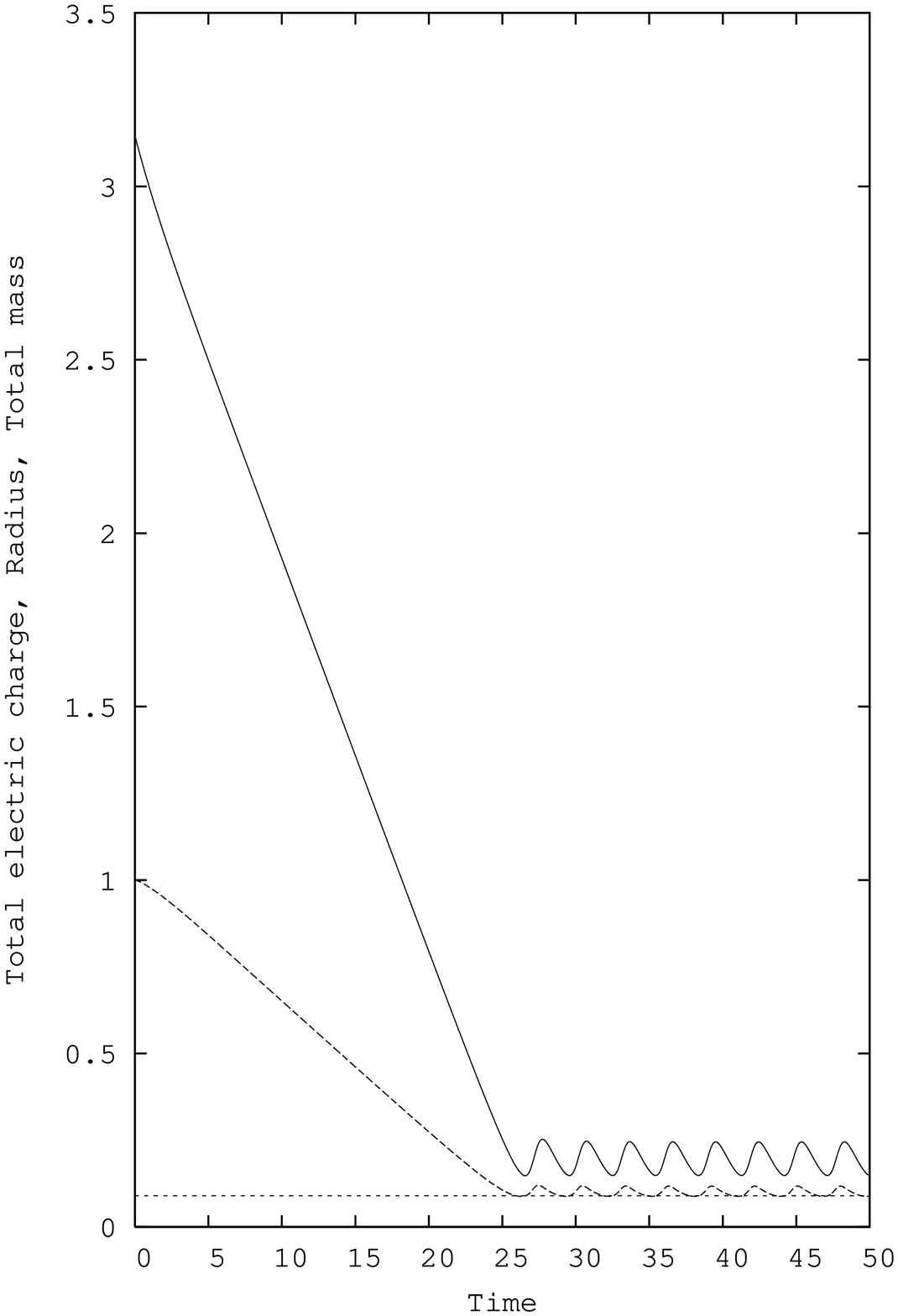}}
\caption{Total electric charge (short dashed line) and the evolution
of radius $a$ (solid line) and the total mass $m_a$ (dashed line).}
\label{one}
\end{figure}

\begin{figure}
\centerline{\epsfxsize=4.in\epsfbox{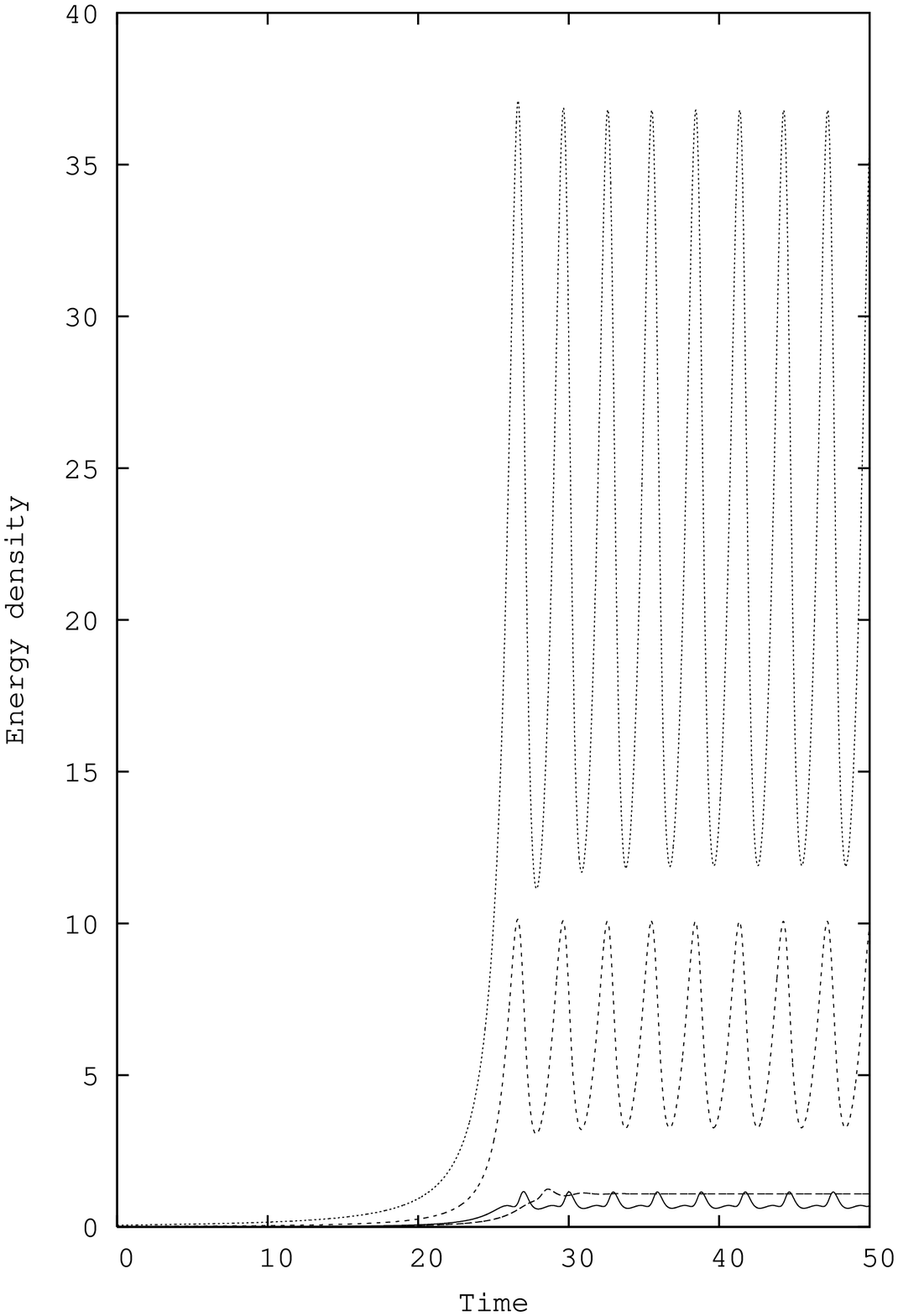}}
\caption{Energy density
 as a function of time for different pieces
of the distribution: $x=0.25$ (short dashed line); $x=0.5$ (large
dashed line); $x=0.75$ (solid line) and $x=1$ (dotted line).}
\label{two}
\end{figure}

\begin{figure}
\centerline{\epsfxsize=4.in\epsfbox{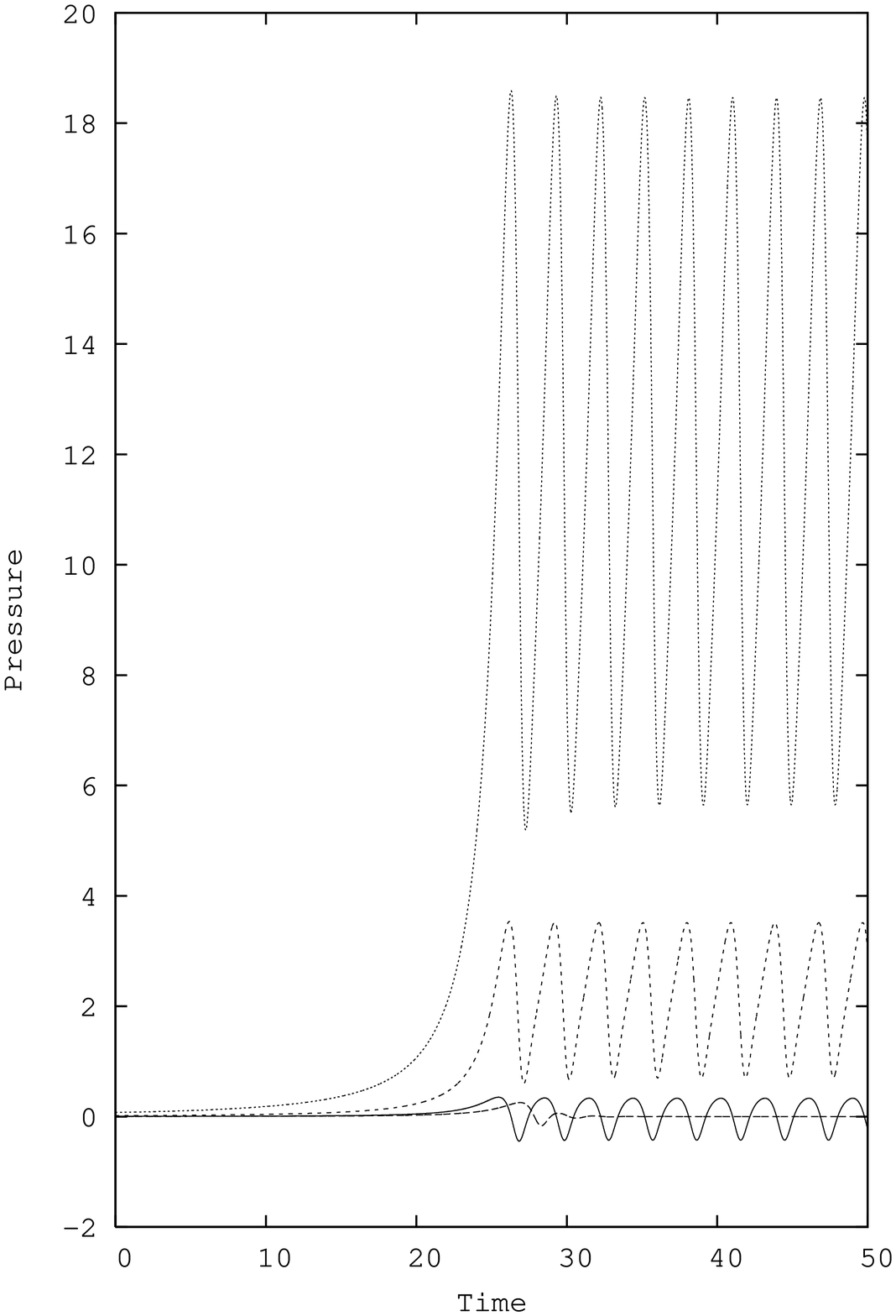}}
\caption{Pressure as a function of time for different pieces
of the distribution: $x=0.25$ (short dashed line); $x=0.5$ (large
dashed line); $x=0.75$ (solid line) and $x=1$ (dotted line).}
\label{three}
\end{figure}

\begin{figure}
\centerline{\epsfxsize=4.in\epsfbox{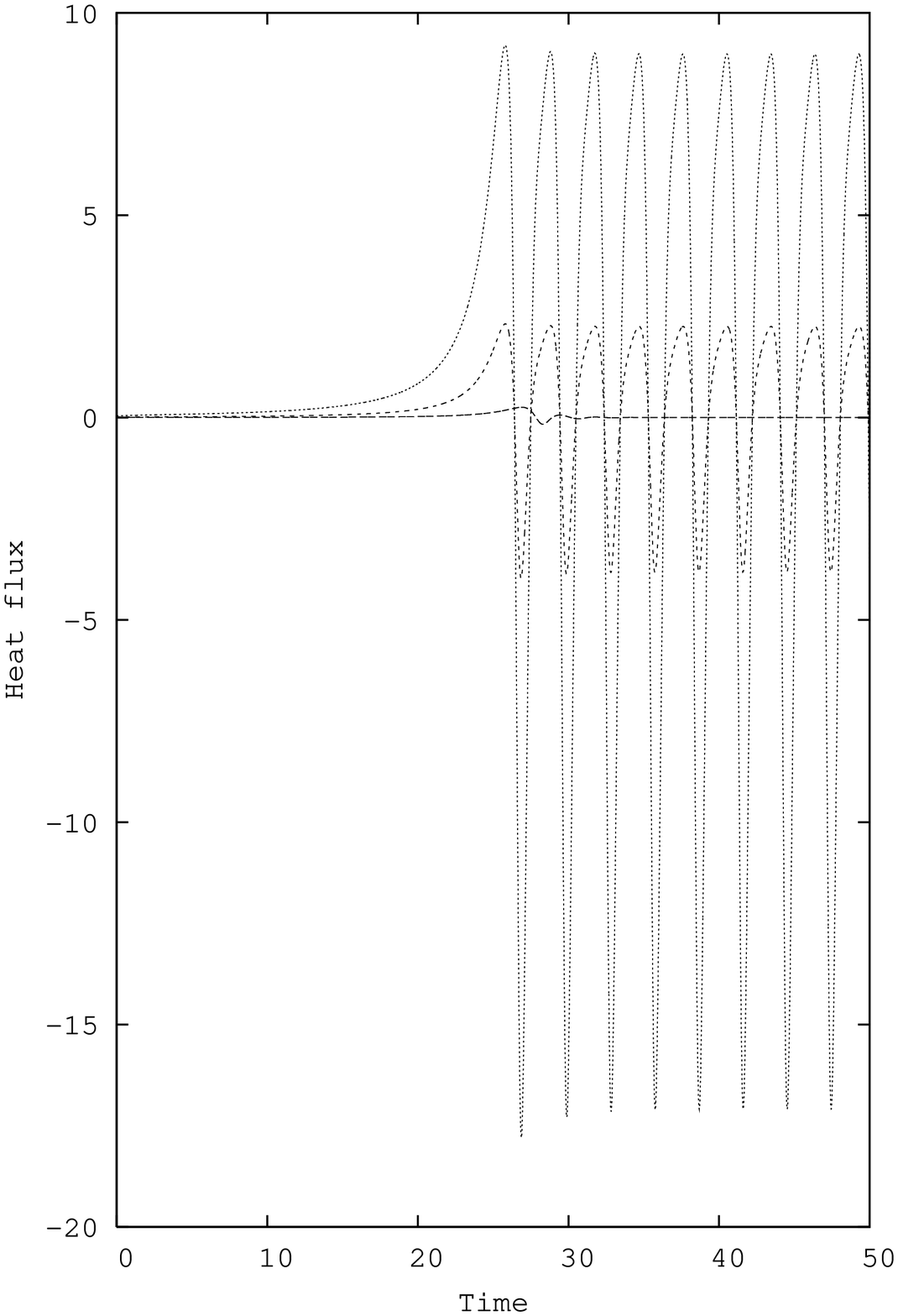}}
\caption{Heat flux as a function of time for different pieces
of the distribution: $x=0.25$ (short dashed line); $x=0.5$ (large
dashed line); $x=0.75$ (solid line) and $x=1$ (dotted line).}
\label{four}
\end{figure}

\begin{figure}
\centerline{\epsfxsize=4.in\epsfbox{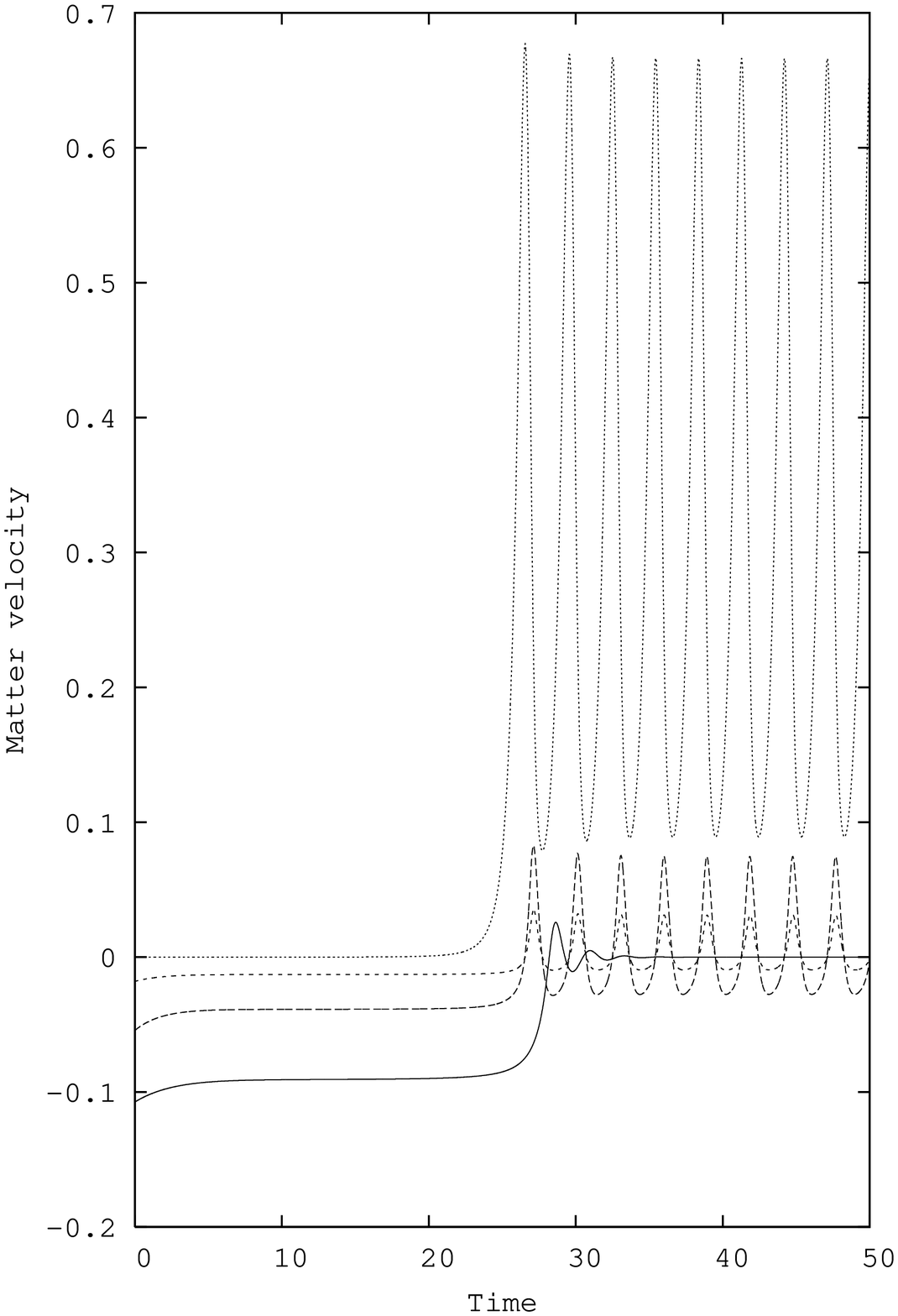}}
\caption{Matter Velocity as a function of time for different pieces
of the distribution: $x=0.25$ (short dashed line); $x=0.5$ (large
dashed line); $x=0.75$ (solid line) and $x=1$ (dotted line).}
\label{five}
\end{figure}

\begin{figure}
\centerline{\epsfxsize=4.in\epsfbox{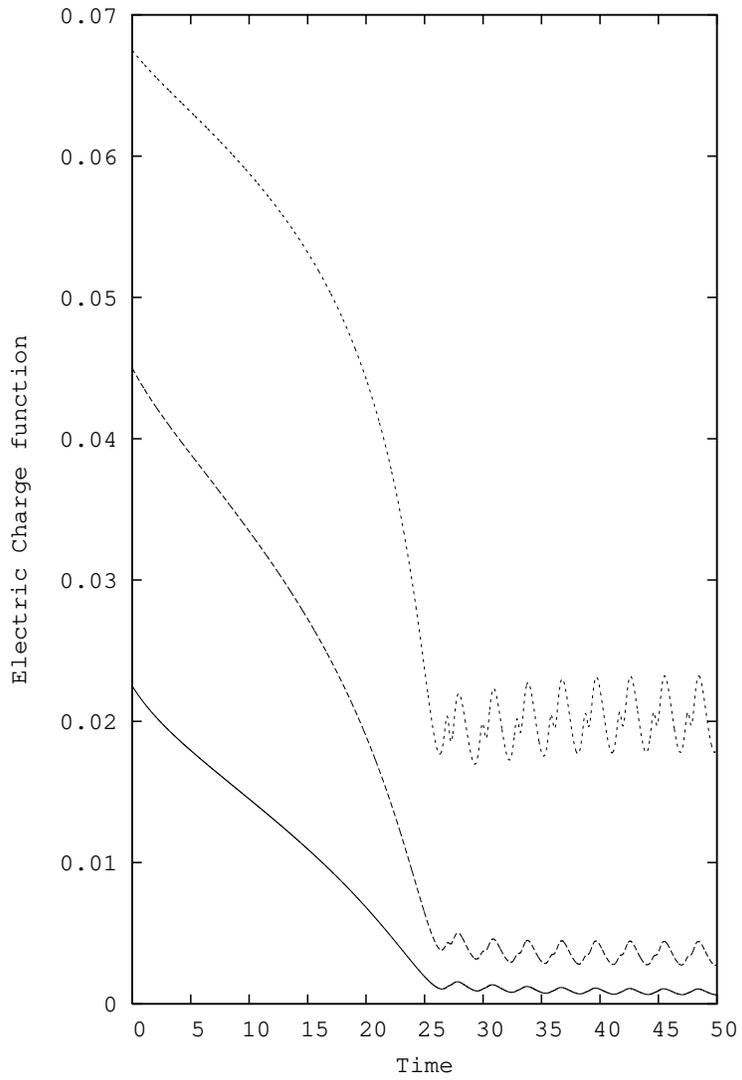}}
\caption{Electric charge function versus time for different
 pieces of the material: $x=0.25$ (solid line); $x=0.5$ (large
dashed line); $x=0.75$ (short dashed line).}
\label{six}
\end{figure}

\begin{figure}
\centerline{\epsfxsize=5.0in\epsfbox{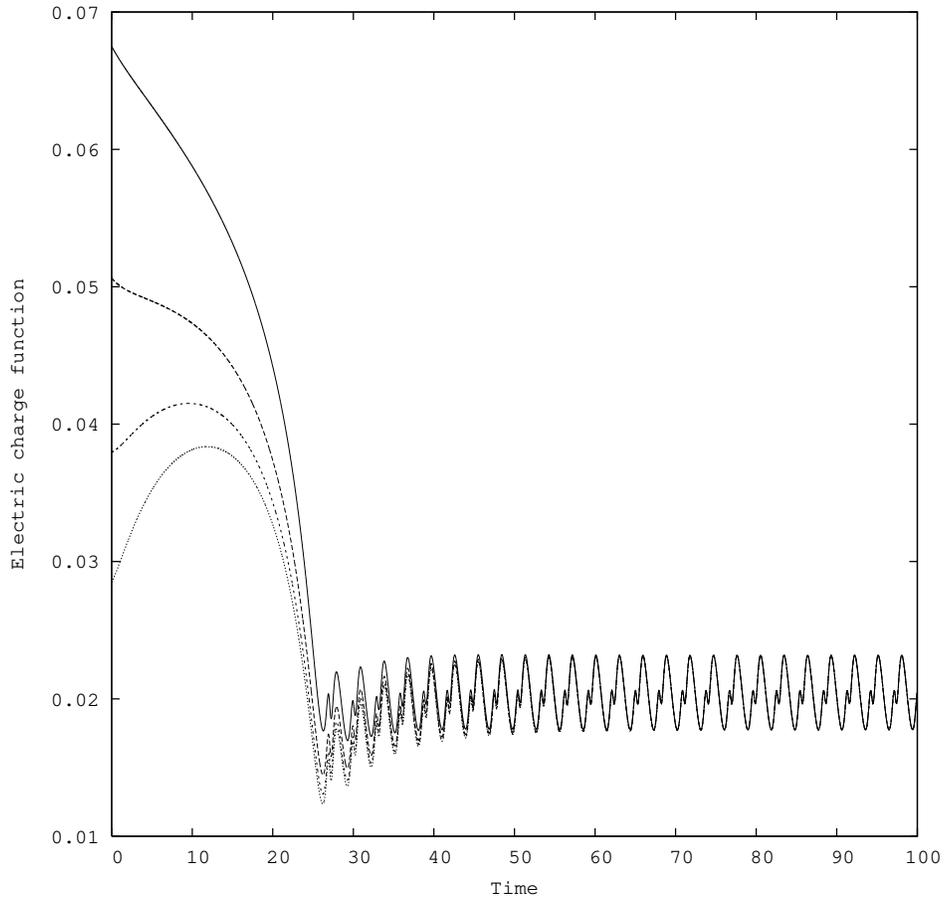}}
\caption{Electric charge function versus time for $x=0.75$ and
different values of power ${\mathcal P}$ 
in the initial value of $C$ to integrate
the conservation equation.  Curves correspond to: ${\mathcal P}=1$
 (continuous line); ${\mathcal P}=2$ (large dashed line);
${\mathcal P}=3$ (short dashed line) and ${\mathcal P}=4$ (dotted line).
Many values of $a(0)$ were used with essentially the
same results.}
\label{seven}
\end{figure}

\begin{figure}
\centerline{\epsfxsize=4.0in\epsfbox{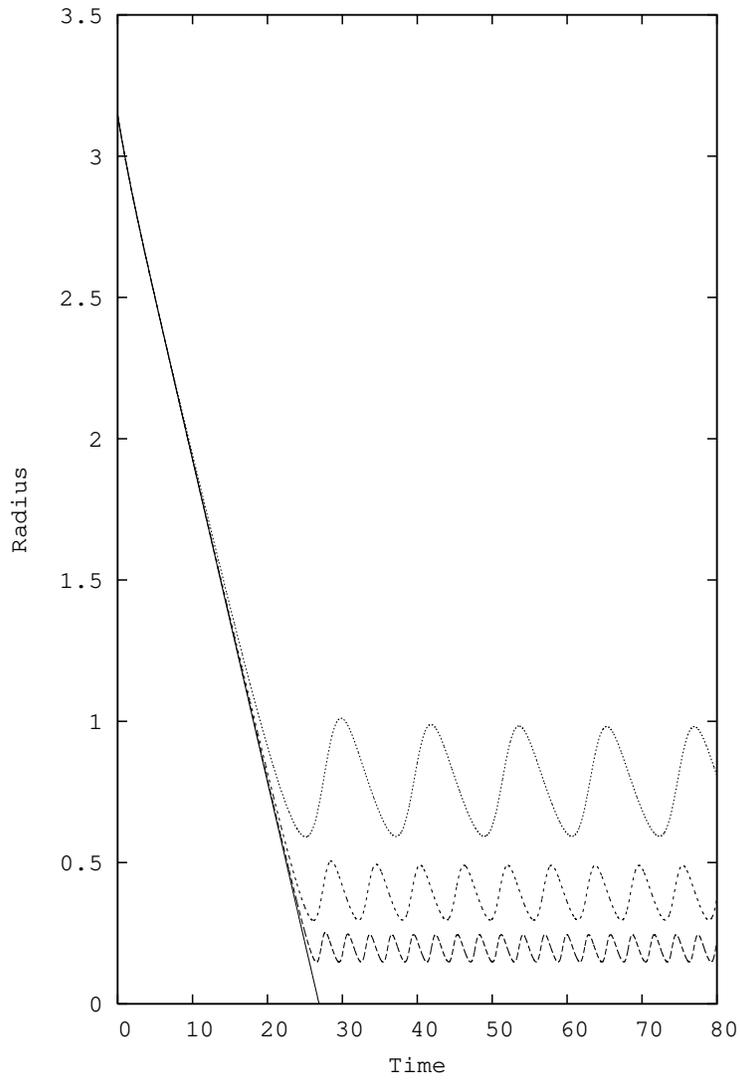}}
\caption{Radius as a function time for $C_T=0.00$ (solid line);
$C_T=0.09$ (large dashed line);
 $C_T=0.18$ (short dashed line); $C_T=0.36$ (dotted line).}
\label{eight}
\end{figure}

Now that we have successfully tested our new approach, we
explore another physical scenario, one with redistribution
of the electric charge. For this case we choose $k=0.1$,
$a(0)=3.5$, $\mu_a(0)=1$ and $C_T=0.09$
to obtain (among other roots) $l=0.741632$.
For ${\mathcal P=1}$ (power of the
initial condition to integrate the conservation equation) the
results are shown in figures \ref{one}--\ref{six}. The system
collapses and losses mass until it reaches a stable oscillatory
state.  When the distribution
is in oscillatory regime the mass behaves in the same way, that is, oscillates between
two extremal values.
This oscillatory behavior
extends to all physical variables at all pieces of the material.
The interior redistribution of electric
charge responds to the surface dynamics.  
Electric charge (repulsively) let the distribution expand with absorption
of energy  up to some maximum point in which gravitation compensates
electric repulsion. Therefore the collapse undergoes with emission of
energy, up to some minimum point in which electric repulsion is again dominant
and so on indefinitely.  This model seems to be independent
 of the initial conditions, as shown in figure \ref{seven}. 
The dependence of the radius on the total electric charge is
shown in figure \ref{eight}. 
Analysis of this last result 
indicates scale invariance with respect to the total electric charge.
In fact, once the stationary state (oscillatory) is reached, the
 amplitude and period depend on the total electric charge.
 Thus, if we know the oscillatory regime for 
some specific value of electric charge, we can reproduce the
oscillatory regime for any other, non zero, value of electric charge.
 In the limit of zero charge the distribution shrink completely,
forming a very small region with very high energy density and pressure.
This last result was found in another context (see section IV. B in Ref. \cite{bpr99}). 

 The energetic at the boundary surface
in conjunction with the pull of gravity and push of electric charge,
explain the dependence between the amplitude and period with
the total electric charge. These features are analogous to the reported
by Brady {\it et al.} \cite{bcg97} for a real and massive scalar field.
In this context the mass destroys the scale invariance
of the Einstein--scalar field equations, but after the nearly critical phase 
the oscillation period depends on the inverse of scalar field mass,
i.e. the Compton wavelength. In our
case the situation is analogous respect to the electric total charge.
Parenthetically, the Seidel--Suen oscillaton \cite{ss91} is the stable solution
far from the critical behavior of a massive and real scalar field.
Our solution behaves like that stable configuration from quite simple
considerations. 

\section{Discussion}\label{conclusion}

In this paper we have shown that electrically charged matter
can be considered as anisotropic matter. 
We exemplify the  approach obtaining 
a dynamical model under the diffusion approximation and 
self--similarity within the source. The showed example
is representative of many others varying only the initial conditions,
$k$ and $C_T$, and imposing the following physical conditions:
$-1<\omega<1$, $p<\rho$ and $\rho> 0$.

Some general considerations concerning our results are  listed next,
\begin{itemize}
\item {\sc Extremely compact distribution:}
     The collapse is halted when the total electric charge is about
      the total mass (in geometrical units). 
      If the initial radius of the electrically charged 
      distribution is $5.2\,\, km$,
       the radius of the final stable source 
      is $\approx 266\,\, m$. Also the final pressure (if not zero)
      is about $ 1.5 \times 10^{39} \,\,Pascal$ and the energy 
      density $ 1.8\times 10^{19}\,\, g/cm^3$.
\item {\sc High electric field:} 
      The electric charge used in our models
      can be as high as $1.5\times 10^{19} \,\,Coulomb $.
      Thus, the electric field for the stable configuration
      is about $2.2\times 10^{19} \,\,V/m$ \cite{b71, remlz03}. 
      The sign of net charge does not affect our results.
\item {\sc High powered:} If the initial total mass is 
  $1 M_{\odot}$ the final is
  $\approx 0.1 M_{\odot}$, that is, almost $90 \%$ of the
  the mass is radiated in $0.15\,\, ms$.
\end{itemize}

A more general solutions to the homothetic
equations (\ref{se1}) and (\ref{se2}) could be necessary to explore
how general our results are. However, the results found using
the anisotropic approach are consistent with previously reported results 
when the static regime is the final state \cite{l95}--\cite{hp85}.
If the solution reaches stationarity it oscillates. For this case 
we did not investigate the stability of the system,
 but independence of initial conditions is apparent.
 It is clear that the presence of some finite electric charge
 (or anisotropy) avoids singularity formation.
Finally, we would like to  conclude by posing a question.
 If the spherically symmetric
source has an upper limit for the total net electric charge
that carries \cite{remlz03,fst03}, is it relaxed by the transport mechanism or
by the additional homothetic symmetry? Work in this direction is in
progress.

\section*{Acknowledgements}
{This work was supported partially by FONACIT 
under grants S1--98003270 and  F2002000426;
by CDCHT--ULA under grant C--1267--04--05--A.
Computer time was provided by the Centro Nacional de
 C\'alculo Cient\'\i fico, Universidad de los Andes (CeCalcULA).}

\newpage


\end{document}